\documentclass[twocolumn,preprintnumbers,amsmath,amssymb]{revtex4}  

\usepackage{graphicx}
\usepackage{dcolumn}
\usepackage{bm}
\usepackage{multirow}
\usepackage{color}
\usepackage{amsmath}
\definecolor{mblue}{rgb}{0,0.35,0.75}

\usepackage[colorlinks]{hyperref}

\begin{document}

\title{Enabling valley selective exciton scattering in monolayer WSe$_2$ through upconversion}

\author{M. Manca$^1$}
\author{M. M. Glazov$^2$}
\email{glazov@coherent.ioffe.ru}
\author{C. Robert$^1$}
\author{F. Cadiz$^1$}
\author{T. Taniguchi$^3$}
\author{K.~Watanabe$^3$}
\author{E. Courtade$^1$}
\author{T. Amand$^1$}
\author{P. Renucci$^1$}
\author{X. Marie$^1$}
\author{G. Wang$^1$}
\email{g\_wang@insa-toulouse.fr}
\author{B. Urbaszek$^1$}
\email{urbaszek@insa-toulouse.fr}

\affiliation{%
$^1$ Universit\'e de Toulouse, INSA-CNRS-UPS, LPCNO, 135 Av. Rangueil, 31077 Toulouse, France\\
$^2$ Ioffe Institute, 194021 St.\,Petersburg, Russia\\
$^3$ National Institute for Materials Science, Tsukuba, Ibaraki 305-0044, Japan}

\begin{abstract}
Excitons, Coulomb bound electron-hole pairs, are composite bosons and their interactions in traditional semiconductors lead to condensation and light amplification. The much stronger Coulomb interaction in transition metal dichalcogenides such as WSe$_2$ monolayers combined with the presence of the valley degree of freedom is expected to provide new opportunities for controlling excitonic effects. But so far the bosonic character of exciton scattering processes remains largely unexplored in these two-dimensional (2D)materials. Here we show that scattering between B-excitons and A-excitons preferably happens within the same valley in momentum space. This leads to power dependent, negative polarization of the hot B-exciton emission. We use a selective upconversion technique for efficient generation of B-excitons in the presence of resonantly excited A-excitons at lower energy, we also observe the excited A-excitons state $2s$. Detuning of the continuous wave, low power laser excitation outside the A-exciton resonance (with a full width at half maximum of 4 meV) results in vanishing upconversion signal.

\end{abstract}
                             
\maketitle
In semiconductors the optical properties are governed by excitons,  bound electron-hole pairs \cite{Gross:1952a,knox1963theory,Rashba:1982a,Kazimierczuk:2014a,Wang:2005a,Klingshirn:2006a}  with certain analogies to the hydrogen atom. The exciton energy states and polarization selection rules need to be understood for designing optoelectronic applications which target efficient emission and strong light-matter interaction \cite{Haug:2004a}. Excitons also provide a rich platform for fundamental physics experiments ~\cite{keldysh68a,moskalenko,Snoke:2002a,Gorbunov:2006aa,sie:2015b}.  A real breakthrough for exciton physics in the solid state was to show that excitons behave like composite bosons \cite{Tassone:1999a,Savvidis:2000a,Butov:2001a}. 
The key effect is stimulated scattering: Bosons preferentially scatter to a quantum state that is already occupied. 
These pioneering works enabled studies into optical amplifiers based on excitons and also new fundamental research on exciton Bose-Einstein condensation, which are still ongoing \cite{Kasprzak:2006a,Amo:2011a,Goblot:2016a,Fraser:2016a,Sanvitto:2016a}.\\
\indent Our target in this work is to look for fingerprints of bosonic interactions in 2D materials. Excitons in transition metal dichalcogenide (TMDC) monolayers (MLs) provide exciting new opportunities for applications and new frontiers in exciton physics for several reasons: (i) with binding energies of several hundred meV \cite{Chernikov:2014a,Zhu:2015b,Ugeda:2014a,Wang:2015b,He:2014a,Hanbicki:2015a,Ye:2014a}, excitons dominate optical properties even at room temperature, (ii) the strong exciton oscillator strength leads to absorption of up to 20 \% per monolayer \cite{Lopez:2013a,Li:2014b}, and (iii) the interband selection rules are valley selective. In combination with strong spin-orbit (SO) splittings this allows studying spin-valley physics \cite{Cao:2012a,Xiao:2012a,Jones:2013a,Mak:2014a,Yang:2015a,Plechinger:2016a}. These unique excitonic properties make ML TMDCs ideal systems for investigating exciton interactions \cite{Mouri:2014a,Kumar:2014b,PhysRevB.93.201111,Sun:2014a,Moody:2015a,Amani:2015a} and microcavity polariton physics \cite{Liu:2015b,Lundt:2016a,Flatten:2016a,Low:2016a,Kolobov2016,Bleu:2016a,Vasilevskiy:2015a,Basov:2016a,Dufferwiel:2015a,Sidler:2016a,Fogler:2014a}.\\
\indent We introduce an original optical excitation scheme to study exciton scattering in the model 2D TMDC monolayer material WSe$_2$. We provide a unique situation for efficient generation of B-excitons in the presence of resonantly excited A-excitons at lower energy. We resonantly excite the A-exciton with a low power laser using an extremely high quality sample with only 4 meV FWHM transition linewidth. Surprisingly, we observe emission from the B-exciton, 430~meV above the A-exciton state and also of the excited A-exciton ($2s$) 130~meV above the fundamental, $1s$, state, which we refer to for brevity as 'upconversion' PL.  For circularly polarized excitation we show that the upconverted B-exciton emission is strongly cross-circularly polarized, the polarization degree increases with laser excitation power. This can be interpreted as a first fingerprint of boson scattering of 2D excitons  \cite{Tassone:1999a,Savvidis:2000a,Butov:2001a} that favours relaxation from the B- to A-excitons within the same valley in momentum space due to the chiral optical selection rules. Possible mechanisms of the upconversion in ML WSe$_2$ are discussed \cite{Jones:2015a} and compared to upconversion reported for more traditional nanostructures such as InP/InAs heterojunctions, CdTe quantum wells and InAs quantum dots \cite{Seidel:1994a,Hellmann:1995a,Poles:1999a,Paskov:2000a,Chen:2012a}. \\
\begin{figure*}[tb]
\includegraphics[width=0.95\textwidth]{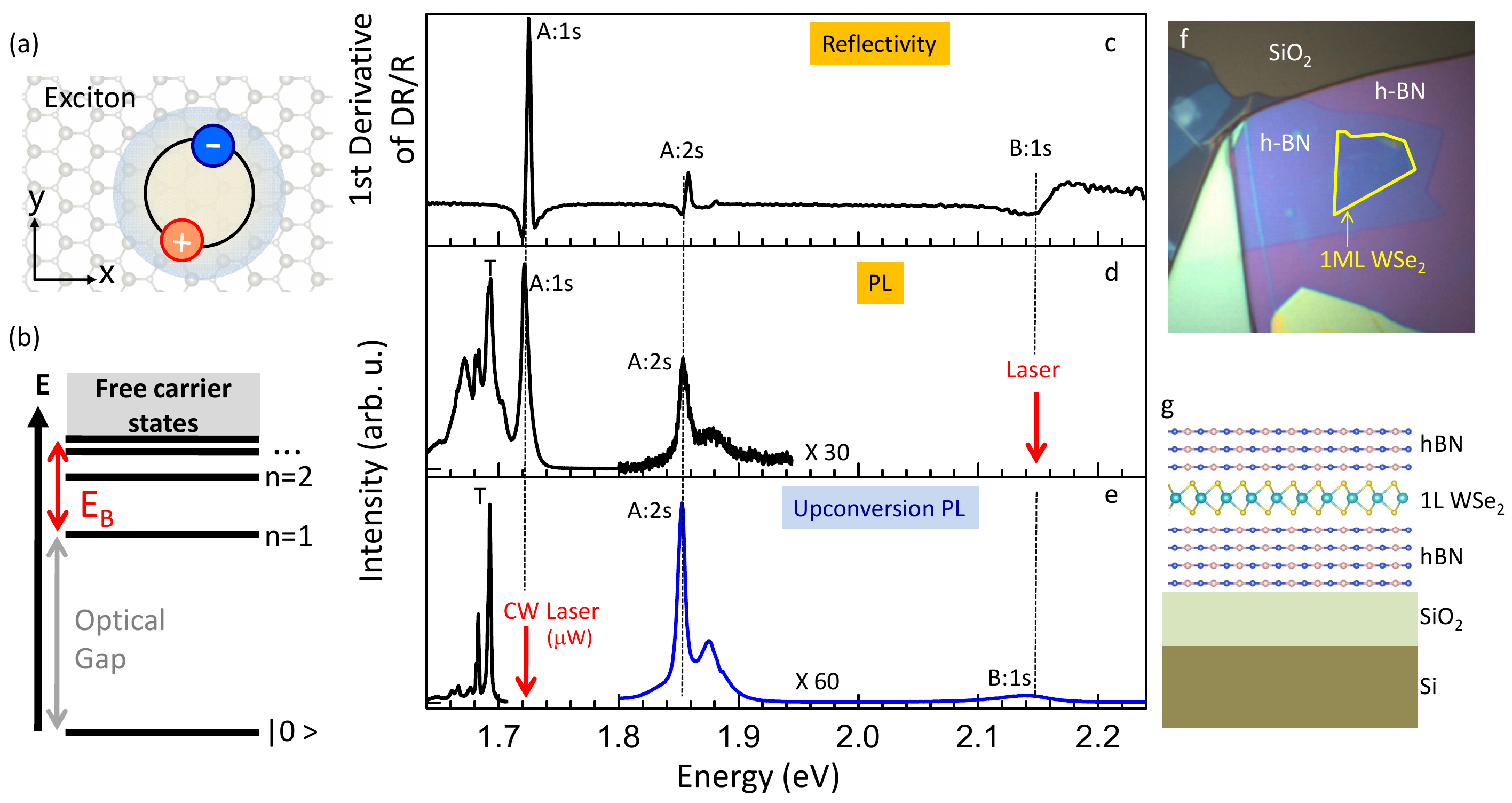}
\caption{\label{fig:fig1} \textbf{Exciton resonances in linear spectroscopy and upconversion at T=4~K}. The sample consists of a WSe$_2$ monolayer encapsulated in hBN \cite{Taniguchi:2007a}.
(a) Strongly bound electron-hole pairs, excitons, dominate optical properties of TMDC monolayers such as WSe$_2$. In WSe$_2$ there exist 2 different exciton series. The B-exciton is about 400 meV above the A-exciton due to spin-obit splitting of the valence band. (b) Excitons have a binding energy $E_B$, defined as the difference between the free particle bandgap and the optical bandgap observed in photoluminescence (PL) emission. $E_B$ is of the order of 500 meV, the first excited state $n=2$ is about 130~meV above the $n=1$ state, marked as A:$2s$ and A:$1s$, respectively, throughout this manuscript. (c) We have performed reflectivity with a white light source to identify the A- and B-exciton at T=4K. In addition we observe an excited states of the A-exciton labelled A:$2s$. (d) In photoluminescence we observe neutral A-exciton and trion emission (T), in addition we see hot luminescence of the A:$2s$ state. (e) We also demonstrate upconversion PL: the laser is tuned to the A:$1s$-exciton resonance and strong emission from B:$1s$ and A:$2s$ at much higher energy is recorded, in addition to the trion emission (T) at lower energy. (f) optical microscope image of the studied van der Waals heterostructure. (g) Scheme of the side view of the sample.
}
\end{figure*}
\begin{figure*}[tb]
\includegraphics[width=0.85\textwidth]{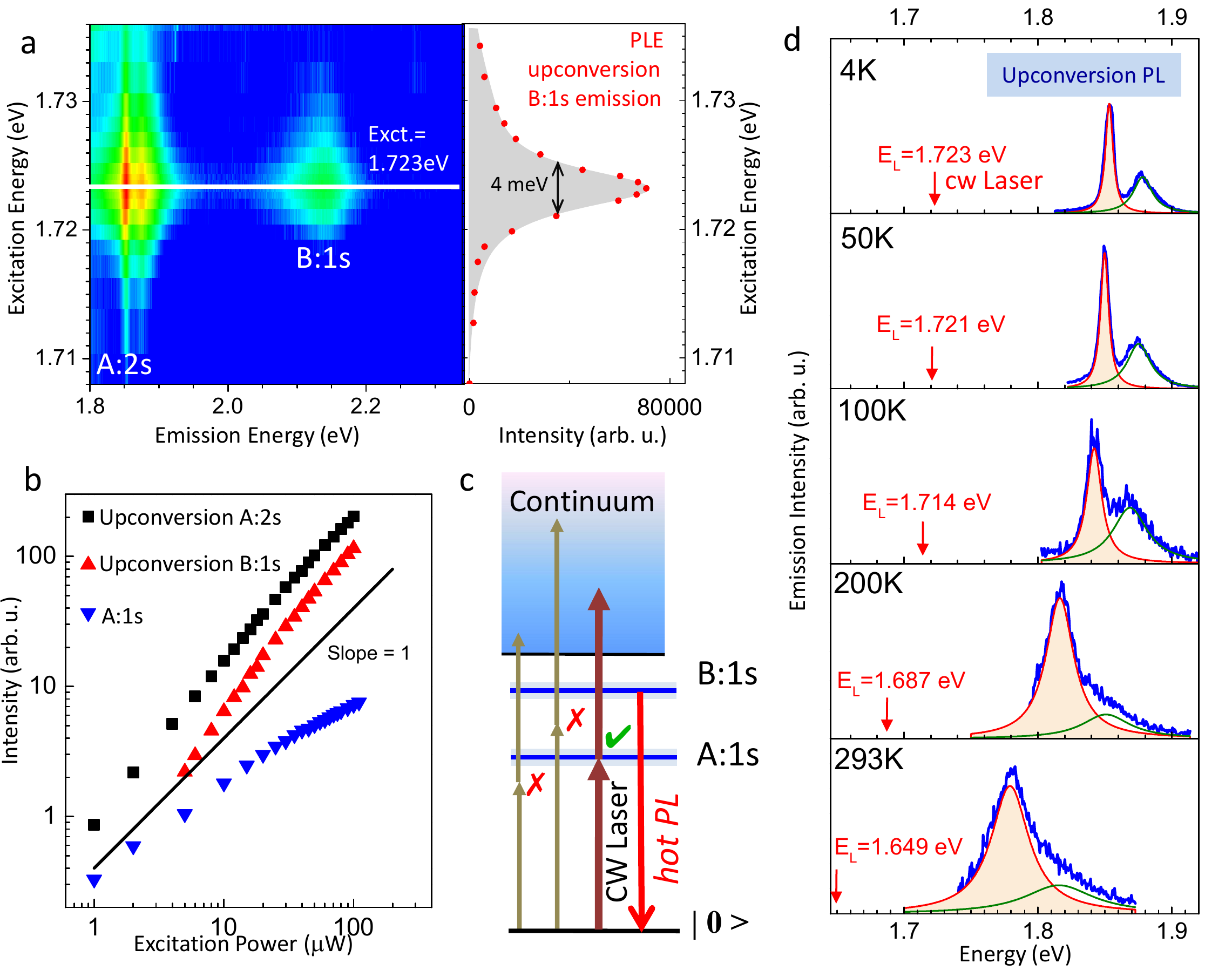}
\caption{\label{fig:fig2} \textbf{Investigating the origin of the upconversion process}. Sample temperature $T= 4$~K. (a) Left panel: contour plot of the upconversion PL intensity for A:$2s$ and B:$1s$ as a function of excitation energy, with a clear resonance at 1.723 eV, the A:$1s$ exciton transition energy. Right panel: The resonance in upconversion PLE has a FWHM of only 4 meV. (b) We show power dependence of the upconversion PL intensity at resonance and compare with power dependence of the neutral exciton excited at the A:$2s$ state in standard PL. Slopes in the range of 10\ldots 100~$\mu$W are for A:$2s$=1.1, B:$1s$=1.25 and A:$1s$=0.56   (c) Scenario for upconversion PL based on 2-photon absorption aided by a real intermediate state is presented. (d) Evolution of upconversion PL of A:$2s$ for $T=4$~K up to room temperature, the excitation laser energy is indicated following the shift of the A:$1s$ exciton resonance with temperature.} 
\end{figure*} 
\begin{figure*}[tb]
\includegraphics[width=0.65\textwidth]{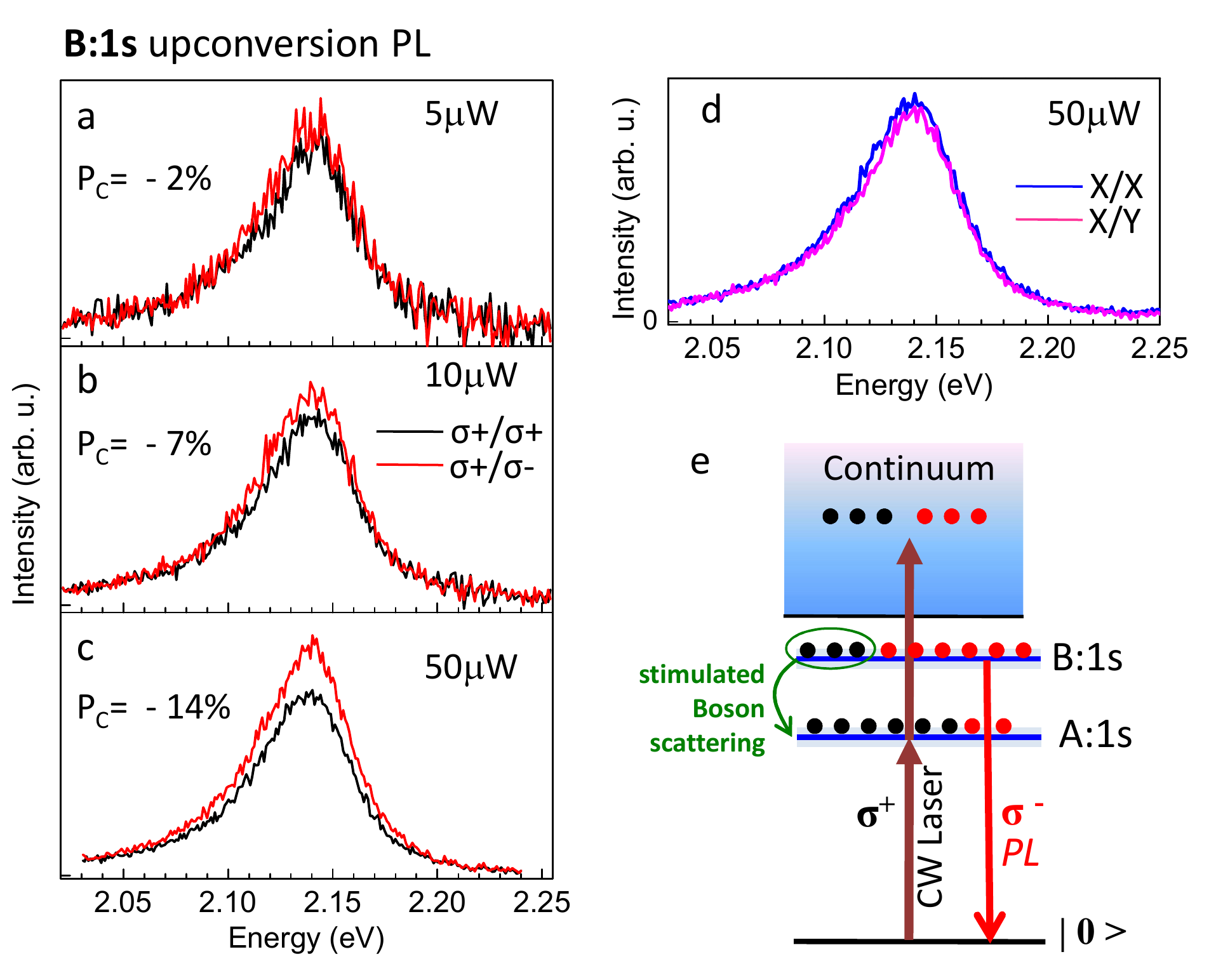}
\caption{\label{fig:fig3}  \textbf{Creating B- and A-excitons: first signatures of boson scattering} T=4~K, $E_L=1.723$~eV, $\sigma^+$ laser polarization. (a) B:$1s$ upconversion PL detected in $\sigma^+$ (black -- \textit{co-polarized}) and $\sigma^-$ (red -- \textit{cross-polarized}) polarization for a laser power of 5 $\mu$W. (b) Same as (a), but for 10 $\mu$W. (c) Same as (a) but for 50 $\mu$W. The value of the circular polarization degree of the emission $P_c=\frac{I^{\sigma+}-I^{\sigma-}}{I^{\sigma+}+I^{\sigma-}}$ is indicated on the panel. As we increase the power, the emission becomes strongly cross-polarized with respect to the initial excitation, we generate a negative polarization $P_c <0$. (d) Laser excitation linearly (X) polarized, detection of upconverted PL at B:$1s$ energy in linear X and Y basis. (e) Scheme to explain negative, power depedent polarization of hot B:$1s$ PL emission observed in panels (a)-(c) based on boson scattering.}
\end{figure*} 

\textbf{RESULTS.}\\
\textbf{Upconversion emission 430~meV above excitation laser.} We study WSe$_2$ MLs encapsulated in hexagonal boron nitride (h-BN) \cite{Taniguchi:2007a}. The aim is to eliminate detrimental surface effects \cite{Amani:2015a} and to provide a symmetric (top/bottom) dielectric environment to study excitons. The high optical quality of the sample is demonstrated in Fig.~\ref{fig:fig1}c: Here we show reflectivity spectra using a white light source for illumination. We detect the A exciton peak at 1.723 eV, with a linewidth of typically 4~meV, this main exciton transition is labelled $1s$ in analogy to the hydrogenic model. About 133~meV above the A:$1s$ we detect an excited exciton state, as demonstrated before for samples exfoliated onto SiO$_2$ \cite{He:2014a,Wang:2015b, Wang:2015g,pollmann:2015a}. The linewidth of this transition, which we tentatively ascribe to the $2s$ A exciton, is of the order of 5 meV. It is accompanied by a smaller peak about 25 meV at higher energy which could be related to the A:$3s$ exciton. At 430 meV above the A:$1s$ we detect the B-exciton transition, where the A-B separation is mainly given by the valence band SO splitting \cite{Kormanyos:2015a}.\\
\begin{figure*}[tb]
\includegraphics[width=0.65\textwidth]{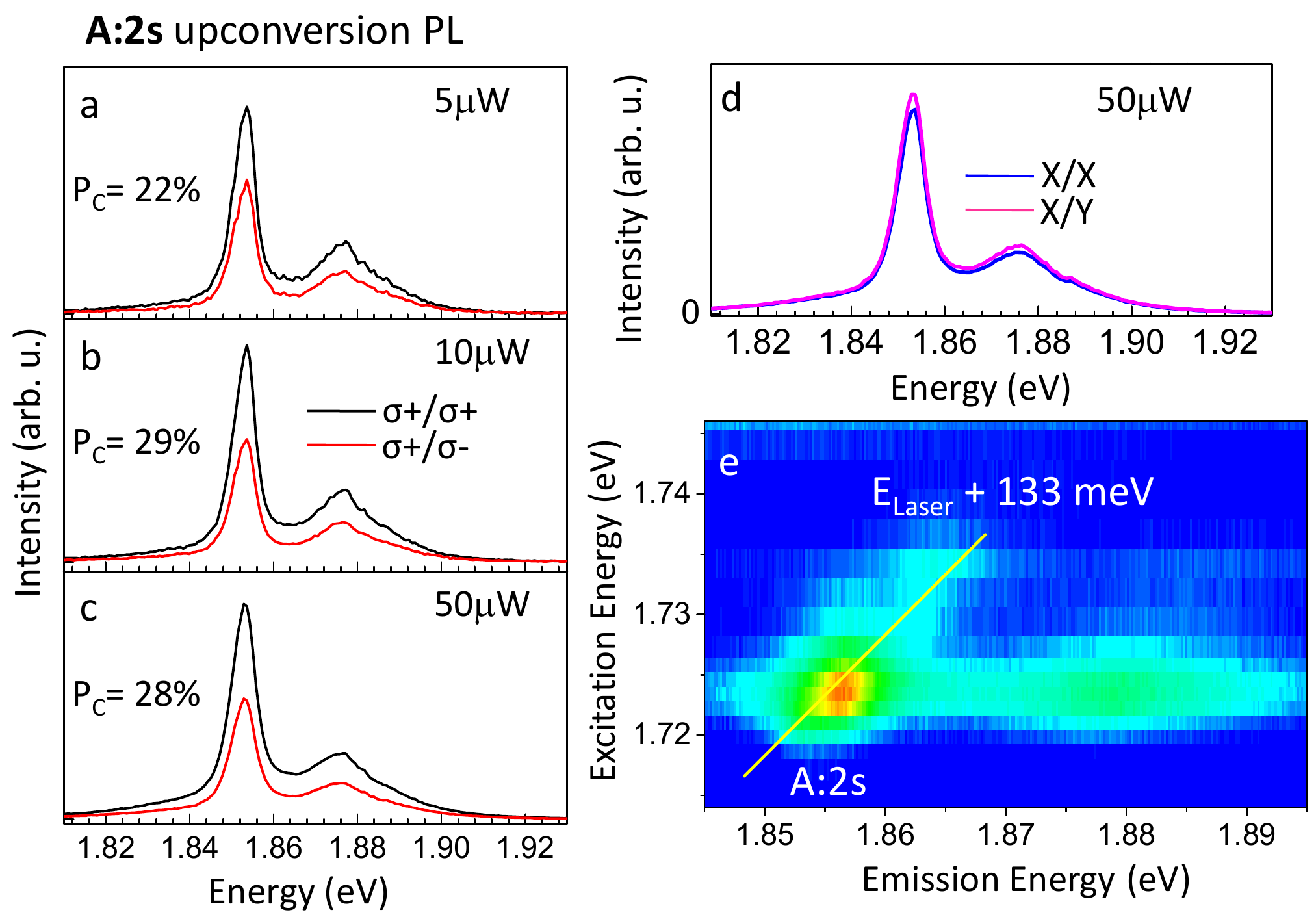}
\caption{\label{fig:fig4}  \textbf{Photon emission 133~meV above the excitation laser at 4 Kelvin.} (a) $T= 4$~K, $E_L=1.723$~eV $\sigma^+$ laser polarization. Upconversion PL of A:$2s$ state detected in $\sigma^+$ (black) and $\sigma^-$ (red) polarization for a laser power of 5 $\mu$W. (b)~Same as (a), but for 10 $\mu$W. (c) Same as (a) but for 50 $\mu$W.  (d) Laser excitation linearly (X) polarized, detection of upconverted PL at A:$2s$ energy in linear X and Y basis. (e) Contour plot of A:$2s$ upconversion PL as a function of laser energy, a Raman feature moving with the laser energy is clearly visible, see supplement for water-fall style plot.}  
\end{figure*} 
\indent In addition to white light reflectivity, we have also performed PL experiments shown in Fig.~\ref{fig:fig1}d. Using an excitation laser energy resonant with the B-exciton transition, we observe in addition to the A:$1s$ exciton emission (4~meV FWHM) and the trion (T) also the hot A:$2s$ emission (5~meV FWHM), at very similar energies as the reflectivity results, indicating negligible Stokes shifts. It is a signature of the high quality of our sample, since absence of the Stokes shifts indicates only very weak localization of excitons, if any. The trion emission is detected in PL but not clearly in reflectivity, which indicates a lower resident electron concentration as compared to the neutral exciton~\cite{astakhov02,Chernikov:2015a}. Hot PL of the A:$2s$ is also observed for other laser energies such as 1.96~eV (HeNe Laser).\\
\indent The optical spectra shown in Fig.~\ref{fig:fig1}e introduce our upconversion scheme: excitation of the A:$1s$ exciton ($E_L=$1.723~eV) results in strong PL emission at \textbf{higher} energy of the A:$2s$ ($E_L+$133~meV) and B:$1s$ exciton ($E_L+$430~meV), as well as trion emission (labelled T) at lower energy $E_L-$31~meV. This upconversion is achieved using a narrow linewidth, continuous wave laser and moderate excitation powers in the $\mu$W$/\mu$m$^2$ range, see Methods. Interestingly we also observe in WSe$_2$ MLs directly exfoliated onto SiO$_2$ (no hBN in the structure) this upconversion emission for A:$2s$ and B:$1s$, see supplement. These observations are very surprising and further experiments that aim to clarify the origin of this upconversion are shown in Fig.~\ref{fig:fig2}.\\

\textbf{Investigating the origin of the upconversion signal.}
The observed upconversion is extremely energy dependent: only strictly resonant excitation of the A:$1s$ exciton results in measurable upconversion luminescence. The FWHM of the observed resonance in upconversion PL excitation (PLE) is about 4 meV, as shown in Fig.~\ref{fig:fig2}a, see red data points. Resonant excitation 31~meV below the A:$1s$ state at the trion energy for example \cite{Jones:2015a}, does not result in emission at the A:$2s$ and B:$1s$ energies in our experiment.  Emission at higher energy than the laser can have several origins \cite{ovsyankin1966mechanism,Seidel:1994a,Hellmann:1995a,Poles:1999a,Paskov:2000a,Chen:2012a}, here important information comes from the B-exciton emission: at 430 meV above the laser energy mechanisms purely based on phonon emission (i.e. laser cooling \cite{Zhang:2016a}) are very unlikely at the sample temperature $T=4$~K. A more probable scenario is 2 photon absorption, made efficient by the A:$1s$ as a \textit{real} intermediate state. This idea is supported by (i) the narrow resonance around the A:$1s$ exciton (Fig.~\ref{fig:fig2}a) and (ii) by analysis of the power dependence of the emission: the upconversion PL evolves with a slope roughly twice that of the exciton A:$1s$ emission, see Fig.~\ref{fig:fig2}c. This indicates that two excitons resonantly excited by the laser combine to form a single excited state of the electron-hole pair, with the energy being the sum of exciton energies. As a plausible scenario we may suggest an Auger-like process, also referred to as exciton annihilation~\cite{Mouri:2014a,Kumar:2014b,PhysRevB.93.201111,Sun:2014a}. In this case the scattering of two existing excitons results in the transition of one electron forming an exciton to the valence band (i.e., nonradiative recombination~\cite{abakumov_perel_yassievich}) while the remaining electron absorbs  the released energy and is promoted to the excited energy band denoted as a continuum $|f\rangle$ in Fig.~\ref{fig:fig2}c. Subsequently, the excited electron-hole pairs loose energy via, e.g., phonon emission and relax towards the radiating states, particularly, B:$1s$ and A:$2s$. As a result, the upconversion intensity scales as $N_{\mathrm{A}:1s}^2$, where $N_{\mathrm A:1s}$ is the exciton occupancy created by the laser. The occupancy $N_{\mathrm A:1s}$ is directly proportional to the intensity of the exciton emission from A:$1s$ state, in agreement with experiment, see also the detailed discussion in the supplement. In addition we observe that upconversion PL in our sample is detectable for the A:$2s$ state even at room temperature, see evolution as a function of temperature in Fig.~\ref{fig:fig2}d.\\

\textbf{Possibility of boson scattering from B- to A-exciton levels}. In our experiment we resonantly pump the A-exciton transition. As addressed above, presumably an additional photon is absorbed to create a second exciton and, ultimately, to generate an electron-hole pair in a high energy continuum state. From there the electron-hole pairs relax towards the B-exciton, where we observe hot luminescence 430~meV above the A-exciton. Very surprisingly the B:$1s$ emissions is strongly $\sigma^-$ polarized following circularly polarized $\sigma^+$ excitation i.e. \textbf{counter-polarized} with respect to the laser, see Fig.~\ref{fig:fig3}a-c.  A strong indication for the importance of scattering processes comes from power dependent measurements: for a laser excitation power of 5 $\mu$W we measure $-3\pm2\%$ PL polarization, for a stronger excitation power of 50 $\mu$W the negative polarization is about $-14\pm2\%$, see panels a-c in Fig.~\ref{fig:fig3}. Observing any polarization at all for these upconversion PL signals is extremely surprising, as it is expected that absorption at high energy continuum states as well as Auger-type processes are not governed by strict polarization selection rules. In this context we have verified that linearly polarized excitation and circularly polarized excitation yield exactly the same intensity of the upconversion signal \cite{Ivchenko:2005a}.\\
\indent How can we explain the strong, negative polarization of the B:$1s$ emission? We propose a scenario based on stimulated boson scattering \cite{Tassone:1999a,Savvidis:2000a,Butov:2001a}, as sketched in Fig.~\ref{fig:fig3}e. As a first step, we assume excitation 
creates an equal population of $\sigma^+$ and $\sigma^-$ excitons in the continuum $|f\rangle$, as chiral selection rules are relaxed. Subsequently the electron-hole pairs relax towards the B:$1s$ state with the same rates for $\sigma^+$ and $\sigma^-$. However, the relaxation from the B-excitons towards the ground, A:$1s$ states is polarization-dependent as bosons preferentially scatter to a quantum state that is already occupied : The pump laser creates a majority of co-polarized A-excitons and since excitons are bosons, the scattering probability of $\sigma^+$/$\sigma^-$ polarized B:$1s$ excitons towards A:$1s$ excitons grows as $(1+N_\text{A:$1s$}^{\sigma^\pm})$~\cite{Butov:2001a}, where $N_{\mathrm{A:}1s}^{\sigma^\pm}$ are the occupancies of correspondingly polarized A:$1s$-excitons. As a result, co-polarized B-exciton states get depleted faster than counter-polarized B-excitons. This imbalance gives rise to hot B-exciton PL emission counter-polarized with respect to  the excitation laser. Linearly polarized excitation of A:$1s$ does not induce any linearly polarized upconversion emission of B:$1s$, see Fig.~\ref{fig:fig3}d. Linear polarization is linked to valley coherence \cite{Jones:2013a}, which is too fragile to be maintained during the upconversion and energy relaxation processes.\\

\textbf{Polarization of A:2s upconversion emission and anti-Stokes Raman at $E_L+133$~meV}. The upconversion PL of the A:$2s$ transition on the other hand is strongly co-polarized with the excitation laser with $P_c\approx 25\%$, here the dependence on excitation laser power is rather weak as shown in Fig.~\ref{fig:fig4}a-c. We argue that the polarization of the A:$2s$ is similar to the A:$1s$ polarization as the exciton populations of the two states are coupled. First, the A:$2s$ to A:$1s$ separation is only 133~meV, compared to the B:$1s$ to A:$1s$ separation of 430 meV. Second, we observe anti-Stokes Raman scattering superimposed on the hot PL of the A:$2s$ exciton, as can be seen in Fig.~\ref{fig:fig4}e and in the supplement. Previously, we have reported double resonant \textbf{Stokes} Raman scattering \cite{Wang:2015g}, that showed efficient relaxation from the A:$2s$ state to the A:$1s$ state as they are separated by a phonon-multiple. Here the equivalent \textbf{anti-Stokes} process is visible in the experiments. Due to efficient phonon exchange between the A:$1s$ and A:$2s$ states the polarizations of the ground and excited states have the same sign.
Note that we cannot probe the A:$1s$ polarization directly in resonant excitation conditions (signal is obscured by scattered laser light). In these experiments at $T=4$~K the phonons can be generated by the relaxation following 2 photon absorption, for example, as well as due to the exciton to trion conversion through phonon emission \cite{Jones:2015a}. Just as for the B:1s upconversion, the experiments using linearly polarized lasers do not result in linearly polarized emission in Fig.\ref{fig:fig4}d. \\
\indent In summary, we demonstrate upconversion photoluminescence in WSe$_2$ monolayers at energies as high as 430~meV above the laser energy. The effect occurs for strictly resonant excitation of the ground A-exciton state $1s$ and is most probably related to 2 photon absorption enabled by a real intermediate state. Very surprisingly, the upconverted PL emission of the B-exciton is counter-circularly polarized with respect to the excitation laser, which provides a fingerprint of stimulated exciton scattering from B- to A-states, which efficiently depletes the co-polarized B-exciton state. We also show upconversion emission at an energy corresponding to the excited state of the A-exciton $2s$. Here strong phonon effects are visible in the form of anti-Stokes emission which exactly shifts with the excitation laser energy $E_L$ as $E_L+133$~meV.

\textbf{METHODS.} \\ 
\textit{Samples.} The WSe$_2$ ML flakes are prepared by micro-mechanical cleavage of a bulk crystal (from 2D Semiconductors) and deposited using a dry-stamping technique on hexagonal boron nitride \cite{Taniguchi:2007a} on SiO$_2$/Si substrates. Subsequently h-BN was deposited on top of the WSe$_2$. Figure~\ref{fig:fig1}d shows an optical microsope image of the fabricated van der Waals heterostructure.\\
\textit{Experimental Set-up.} Low temperature PL and reflectance measurements were performed in a home build micro-spectroscopy set-up build around a closed-cycle, low vibration attoDry cryostat with a temperature controller ($T=4$~K to 300~K). For PL at a fixed wavelength of 633 nm a HeNe laser was used, for PL experiments as a function of excitation laser wavelength we used a tunable, continuous wave Ti-Sa Laser SolsTis from M~SQUARED allowing continuous tuning in the range of 700-1000 nm. For wavelength below 700 nm the sample is excited by picosecond pulses generated by a tunable frequency-doubled optical parametric oscillator (OPO) synchronously pumped by a mode-locked Ti:Sa laser. The typical pulse and spectral width are 1.6 ps and 3 meV, respectively; the repetition rate is 80 MHz \cite{Lagarde:2014a}. The white light source for reflectivity is a halogen lamp with a stabilized power supply. The emitted and/or reflected light was dispersed in a spectrometer and detected by a Si-CCD camera. The excitation/detection spot diameter is $\approx1\mu$m, i.e. smaller than the typical ML diameter. \\


\noindent\textbf{Acknowledgements.---} We thank ERC Grant No. 306719, ITN Spin-NANO Marie Sklodowska-Curie grant agreement No 676108 , ANR MoS2ValleyControl, Labex NEXT project Bi-3 and LIA CNRS-Ioffe RAS ILNACS for financial support. X.M. also acknowledges the Institut Universitaire de France. M.M.G. is grateful to RFBR, Russian Federation President grant and Dynasty Foundation for partial support. K.W. and T.T. acknowledge support from the Elemental Strategy Initiative
conducted by the MEXT, Japan and JSPS KAKENHI Grant Numbers JP26248061,JP15K21722
and JP25106006. We thank Alexey Chernikov and E.L. Ivchenko for very fruitful discussions. \\

\noindent\textbf{Author Contributions.---} M.M., G.W., F.C. and E.C. performed the measurements. K.W. and T.T. grew the hBN.  C.R. and P.R fabricated and tested the van der Waals heterostructures. M.M.G, T.A., X.M. and B.U interpreted the data, B.U., G.W. and M.M.G wrote the manuscript with input from all the authors.\\

\textbf{Additional information} Competing financial interests: The authors declare no competing financial interests.

\end{document}


\title{Enabling valley selective exciton scattering in monolayer WSe$_2$ through upconversion: \\
\textit{Supplementary Information}}

\author{M. Manca$^1$}
\author{M. M. Glazov$^2$}
\email{glazov@coherent.ioffe.ru}
\author{C. Robert$^1$}
\author{F. Cadiz$^1$}
\author{T. Taniguchi$^3$}
\author{K. Watanabe$^3$}
\author{E. Courtade$^1$}
\author{T. Amand$^1$}
\author{P. Renucci$^1$}
\author{X. Marie$^1$}
\author{G. Wang$^1$}
\email{g\_wang@insa-toulouse.fr}
\author{B. Urbaszek$^1$}
\email{urbaszek@insa-toulouse.fr}

\affiliation{%
$^1$ Universit\'e de Toulouse, INSA-CNRS-UPS, LPCNO, 135 Av. Rangueil, 31077 Toulouse, France\\
$^2$ Ioffe Institute, 194021 St.\,Petersburg, Russia\\
$^3$ Advanced Materials Laboratory, National Institute for Materials Science, Tsukuba, Ibaraki 305-0044, Japan}

\maketitle

\textbf{EXPERIMENTS}

\begin{figure}[tb]
\includegraphics[width=\linewidth]{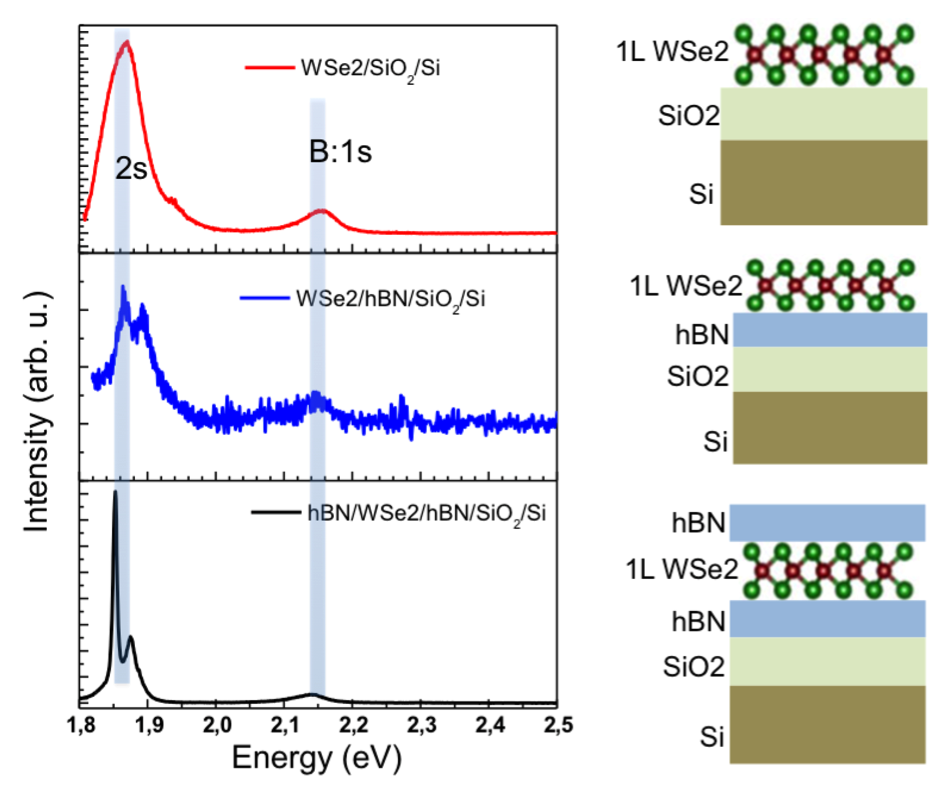}
\caption{\label{fig:fig4}  In the main text the upconversion and its polarization dependence is studied in detail for hBN / 1ML WSe$_2$ / hBN (bottom panel) because of the spectrally sharp and intense exciton transitions. But this effects is not linked to encapsulation, very strong upconversion when exciting the A:$1s$ resonantly is also observed for vacuum / 1ML WSe$_2$ / SiO$_2$ (top panel), and vacuum / 1ML WSe$_2$ / hBN (middle panel).} 
\end{figure}

\begin{figure}[tb]
\includegraphics[width=\linewidth]{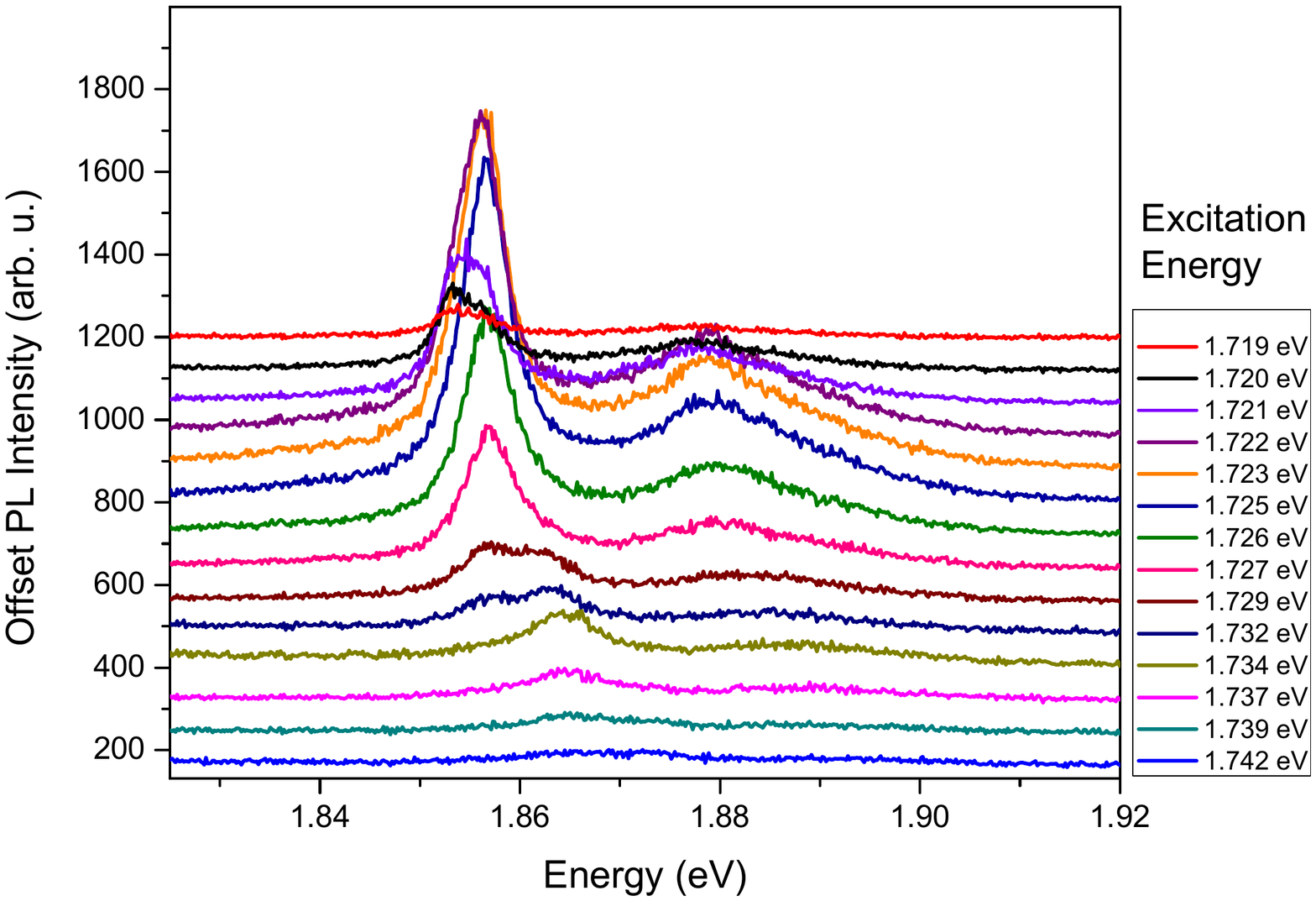}
\caption{\label{fig:fig6}  Upconversion PL of A:$2s$ state as a function of excitation laser energy. The laser energy $E_L$ is indicated for each spectrum. The data is the same as in Fig.4e in the main text, just plotted in a different format.} 
\end{figure}

\begin{figure}[tb]
\includegraphics[width=\linewidth]{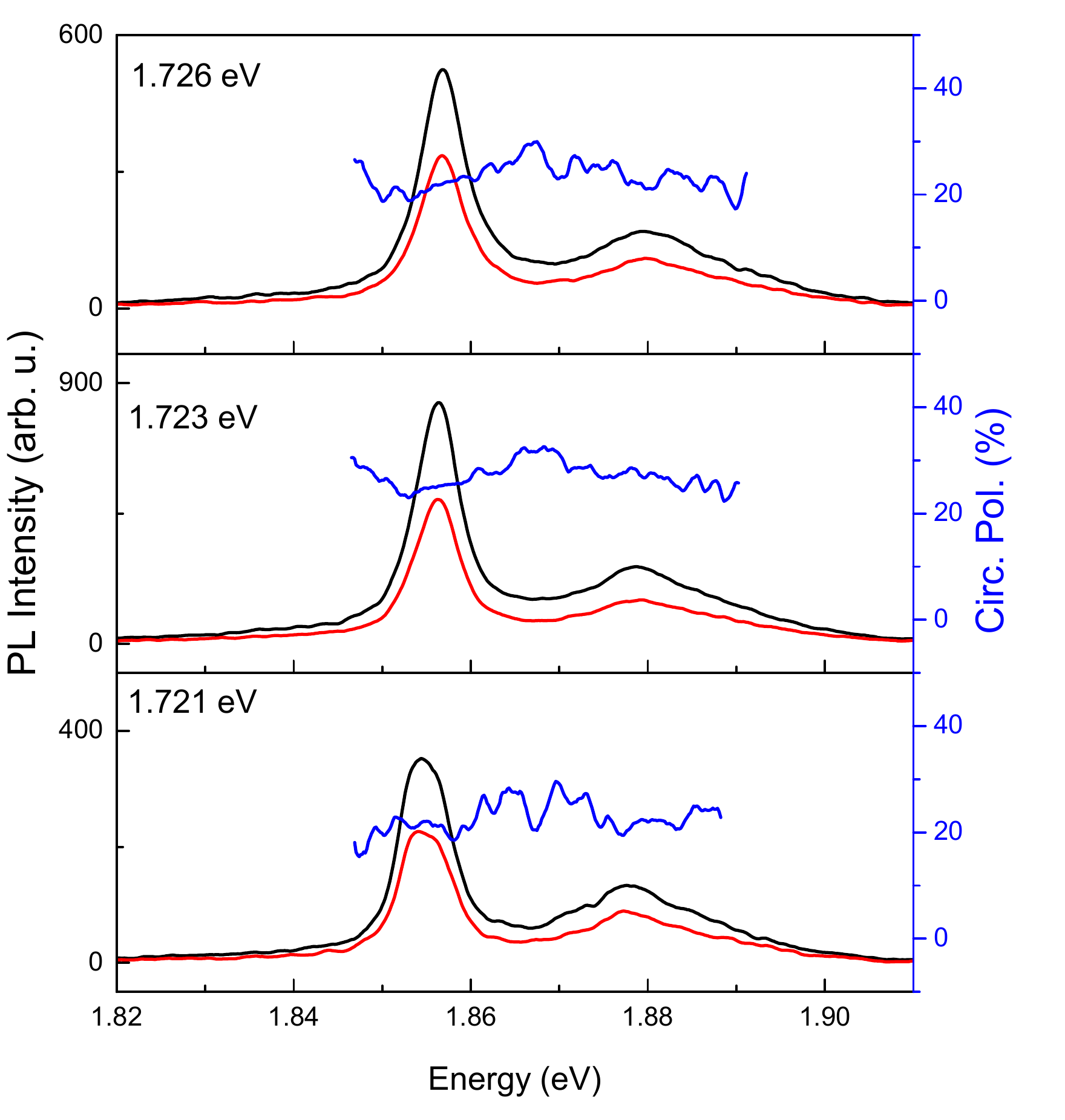}
\caption{\label{fig:fig5} Circulalry polarized excitation, co-circular detection (black) and cross-circular (red) detection of A:$2s$ upconversion PL at T=4K signal (left axis). The circular polarization degree is plotted in blue (right axis). The laser energy $E_L$ is indicated for each panel. } 
\end{figure}

In the main text  upconversion and its polarization dependence is studied in detail for ML WSe$_2$ encapsulated in hexagonal boron nitride (hBN) \cite{Taniguchi:2007a}. This hBN / 1ML WSe$_2$ / hBN has been chosen for its the spectrally sharp and intense exciton transitions, see Fig.~\ref{fig:fig4} bottom panel. But the reported effects are not only observed in encapsulated samples. We also observe very strong upconversion when exciting the A:$1s$ resonantly in standard WSe$_2$ monolayers directly exfoliated onto SiO$_2$, see Fig.~\ref{fig:fig4} top panel and also in uncovered samples 1ML WSe$_2$ / hBN (middle panel).\\
In addition to the A:$2s$ hot PL emission from upconversion we also observe anti-Stokes Raman scattering. This can be seen in Fig.~\ref{fig:fig6}, where in addition to the A:$2s$ a peak 133~meV above the laser energy is clearly visible in each spectrum, shifting as a function of the laser energy $E_L$. This is the same data as in Fig.~4e in the main text.\\
In Fig.~\ref{fig:fig5} we plot  circularly co- and cross-polarized emission in black and red, respectively, of the A:$2s$ upconversion PL. The circular polarization degree $P_c$ is plotted in blue. The polarization of the upconverted emission at the A:2s energy does not originate exclusively from the anti-stokes Raman process, the emission is globally polarized, not just at the Raman energy $E_L+133$~meV. The A:$1s$ resonance is at 1.723 eV, so the top (bottom) panel shows excitation 3~meV above (2~meV below) resonance. The middle panel corresponds to resonant A:$1s$ excitation.

\textbf{MODELS}

\textbf{One-photon absorption.} In linear absorption one absorbed photon generates one exciton. Hence, for resonant excitation of the A:$1s$ state the exciton occupancy $N_{\mathrm{A}:1s}$ is directly proportional to the light intensity~$I$. In the linear regime the exciton generation rate can be conveniently presented as~\cite{Seidel:1994a}
\begin{equation}
\label{1PA}
G = \mathcal A(1-\mathcal R) \frac{I}{\hbar\omega}.
\end{equation}
Here $\mathcal A$ is the absorption coefficient of the TMD ML and $\mathcal R$ is the reflection coefficient of the sample, $\hbar\omega$ is the photon energy. Due to nonlinear effects, e.g., absorption saturation, or nonlinear (Auger) recombination of excitons the exciton occupancy can be sublinear function of the incident light intensity. The exciton occupancy 
\[
N_{\mathrm{A}:1s} \propto G \tau_A,
\]
where $\tau_A$ is the lifetime of the A-exciton. The nonlinearities are either included in the intensity dependence of the absorption coefficient $\mathcal A = \mathcal A(I)$ or in the exciton occupancy dependence of exciton lifetime, $\tau_A=\tau_A(N_{\mathrm{A}:1s})$. The analysis of the particular origin of the sublinear behavior of $N_{\mathrm{A}:1s}$ vs. intensity seen in Fig.~2b of the main text is beyond the scope of the present work.

\textbf{Two-photon absorption and Auger-like process.} The observation of B:$1s$ exciton emission at A:$1s$ excitation is possible if any only if, in addition to the photogenerated exciton, another quantum: second exciton, photon or phonon is involved. Otherwise the energy conservation law is violated. Let us consider in more detail the processes without phonon involvement. 

Standard two-photon absorption (2PA) involves a \textit{virtual} intermediate state. In this case the generation rate of the excitons in the highly excited states denoted as $|f\rangle$  in Fig.~2c of the main text (energy equals to $2\hbar\omega$, i.e. twice the laser energy) is proportional to the square of incident radiation intensity~\cite{ivchenko:2005a,PhysRevB.92.085413,2016arXiv161006780G}
\begin{equation}
\label{2PA}
G_f^{\mathrm{2PA}} \propto I^2.
\end{equation}

\begin{figure}[tb]
\includegraphics[width=0.45\linewidth]{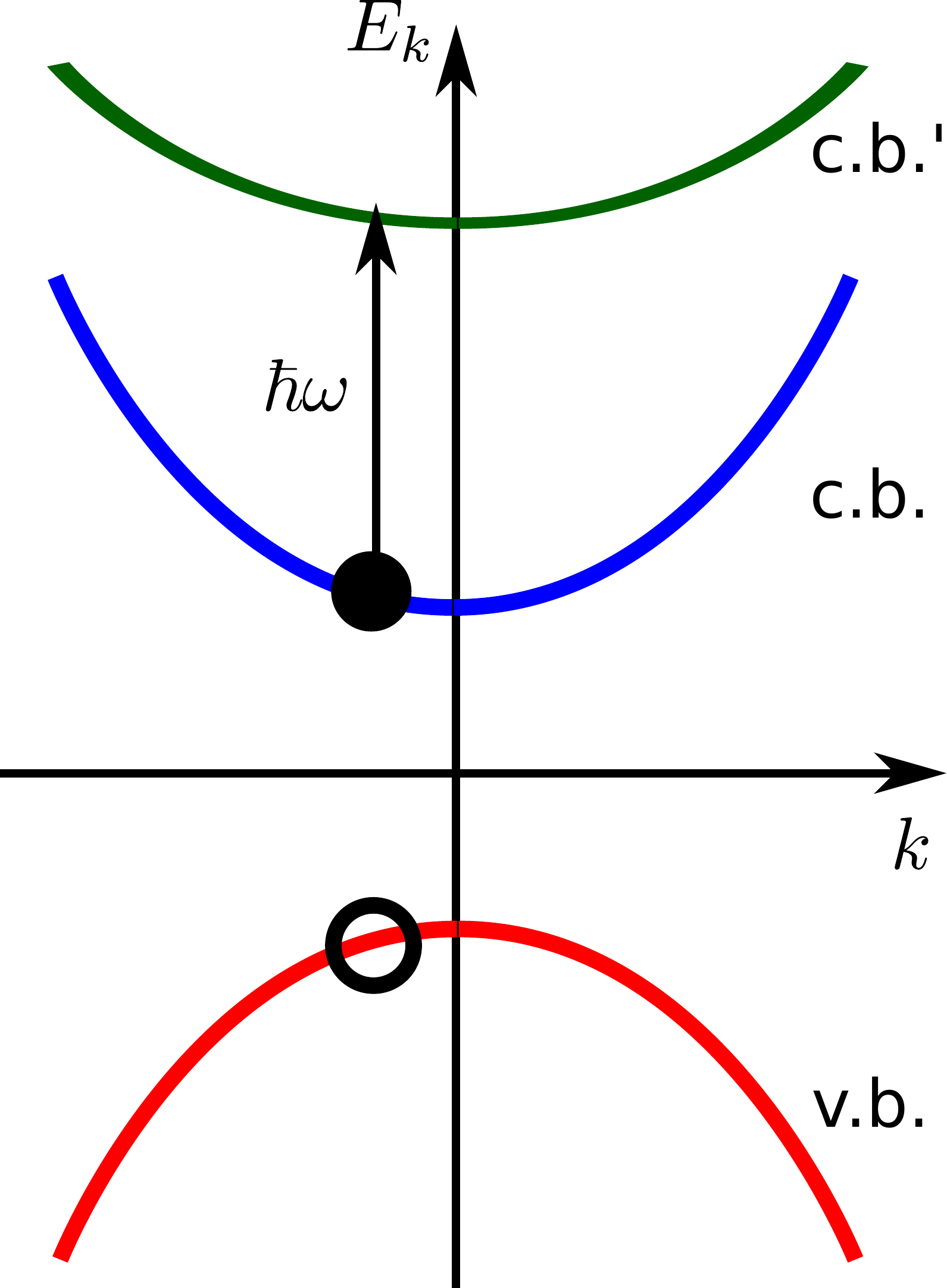}
\caption{\label{fig:RS-2PA} Sketch of the two-photon excitation with a real intermediate state. Open circle denotes the empty state in the valence band (v.b.), filled circle denotes occupied state in the conduction band (c.b.). The arrow denotes the transition to the final state (in the remote conduction band c.b.') through absorption of a second photon. The Coulomb interaction induces spread of the electron-hole pair in $\bm k$-space (not shown) relaxing the momentum conservation law. The band gaps and dispersion are not shown to scale.} 
\end{figure} 

In the studied situation the single photon energy $\hbar\omega$ equals to the A:$1s$-exciton energy. Hence, the intermediate state for the two-photon process can be real. The two-photon excitation via real state (RS-2PA) can be viewed as a two-step process: creation of the A-exciton at the first step and transition of the A-exciton to the excited $|f\rangle$ state via the second photon absorption. On the level of free electron-hole pairs the process is illustrated in Fig.~\ref{fig:RS-2PA}. In this process the generation rate of the excited excitons takes the form
\begin{equation}
\label{2PA}
G_f^{\mathrm{RS-2PA}} \propto N_{\mathrm{A}:1s} I.
\end{equation}
This dependence is weaker than $I^2$ due to possible saturation of linear absorption. Particularly, if intermediate states saturate $G_f^{\mathrm{RS-2PA}} \propto I$.

\begin{figure}[tb]
\includegraphics[width=\linewidth]{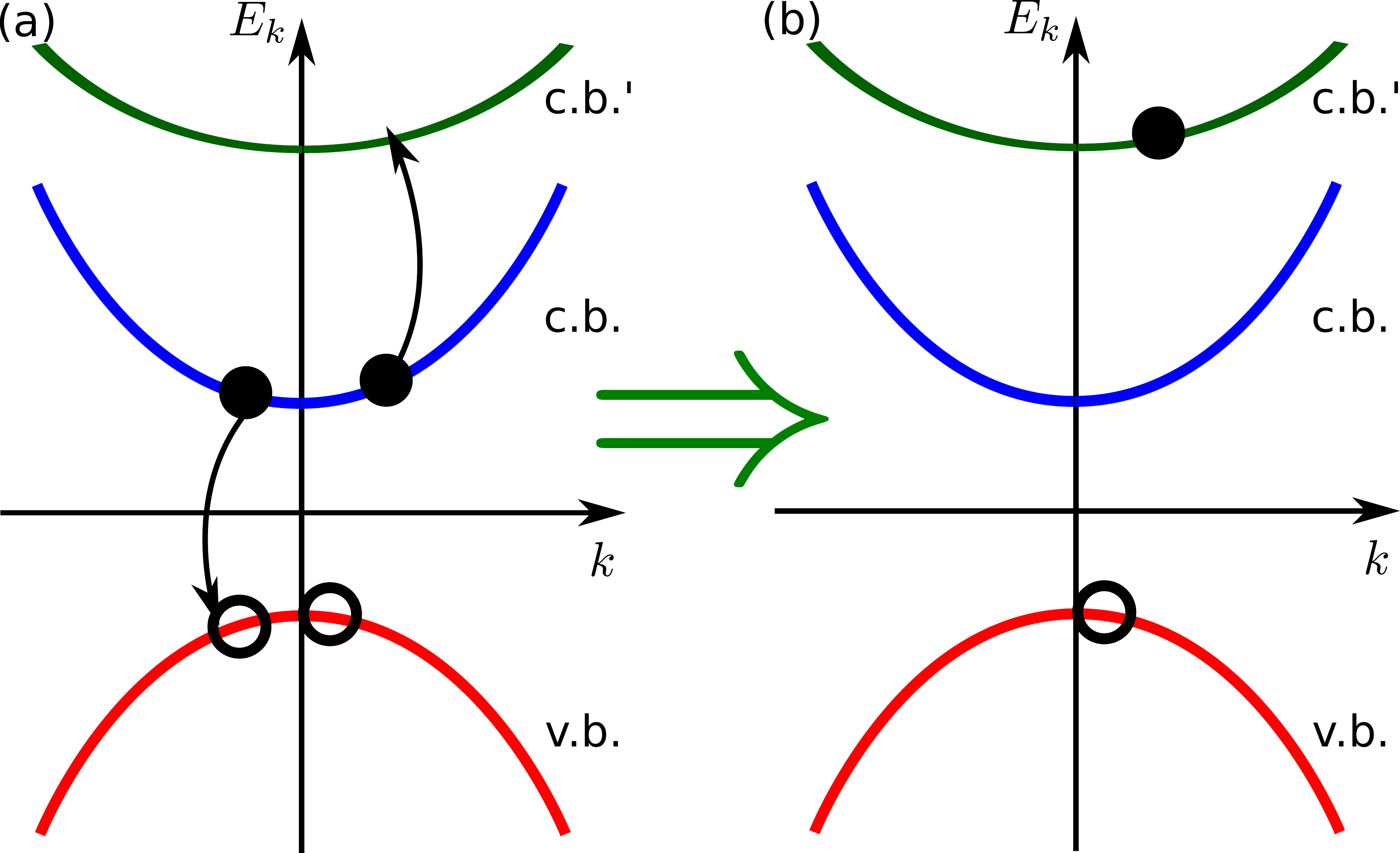}
\caption{\label{fig:Auger} Sketch of the Auger-like two-exciton process. (a) Initial ``two A-exciton state'' formed by independent absorption of two photons. Arrows denote the transitions induced by the Coulomb interaction between electrons. (b) Final state of the electron-hole pair with electron occupying excited band (c.b.'). Other notations are the same as in Fig.~\ref{fig:RS-2PA}} 
\end{figure} 

\begin{table*}[htb]
\caption{Comparison of upconversion intensities for various phonon-less mechanisms: 2PA is the two-photon absorption via virtual state, RS-2PA is the two-photon absorption involving real intermediate states, Fig.~\ref{fig:RS-2PA}, ``sat'' denotes saturation of the real intermediate state, Auger is the Auger-like two-exciton process, Fig.~\ref{fig:Auger}.}
\begin{tabular}{ x{3cm}| x{2cm}|x{2cm}|x{5cm}|x{2cm}}  
Process & 2PA & RS-2PA & RS-2PA (sat) & Auger \\ 
\hline
Intermediate & virtual & real & real & \\
state & & & intermediate states saturate & \\
\hline
Generation rate& $\propto I^2$ & $\propto N_{\mathrm{A}:1s}I$ & $\propto N_{\mathrm{A}:1s}^{(\mathrm{sat})}I(\propto I)$ & $\propto N_{\mathrm{A}:1s}^2$
\end{tabular}\label{tab:intens}
\end{table*} 

Moreover, two photons can be absorbed independently resulting in formation of two A:$1s$-excitons as schematically shown in Fig.~\ref{fig:Auger}a. The Coulomb interaction between the charge carriers forming the excitons results in the redistribution of the excitation in the $\bm k$- and energy-spaces. Particularly, the Auger-like process is possible. In this scenario due to the Coulomb scattering one of the interacting electrons goes to the unoccupied state in the valence band, while another one takes the released energy and gets promoted to the excited energy band~\cite{abakumov_perel_yassievich}. As a result, the excited $|f\rangle$ state of the electron-hole pair is formed, Fig.~\ref{fig:Auger}b. The strong Coulomb interaction between the electron and the hole results in the spread of excitonic functions in the $\bm k$-space and relaxes the momentum conservation law. For the Auger-like process the generation rate of excitons in the excited states is quadratic in the occupancy of A:$1s$ excitons
\begin{equation}
\label{2PA}
G_f^{\mathrm{Auger}} \propto N_{\mathrm{A}:1s}^2.
\end{equation}

Table~\ref{tab:intens} summarizes the results of the analysis performed above on the dependence of the exciton generation rate on the incident laser intensity. We assume that the relaxation from $|f\rangle$ exciton states towards A:$2s$- and B:$1s$-excitons is linear,i.e., its rate is proportional to the first power of the $|f\rangle$ states occupancy. Therefore, the upconversion intensity is proportional to the generation rate of the excited excitons, $G_f$. The comparison of the experimentally observed upconversion PL intensity as a function of the laser power, Fig. 2b of the main text, with the suggested mechanisms shows that the Auger-like process is the plausible source of the upconversion. The 2PA via real states is also possible if the substantial saturation of the intermediate states is assumed. However, the rise of the temperature up to the room temperature keeps upconversion efficient, Fig.~2d of the main text. This rules out trap states, which usually play a role of intermediate states for RS-2PA in the quantum well structures~\cite{Hellmann:1995a}, as the origin of the intermediate states here.

\textbf{Bose stimulation in the relaxation process.} In order to illustrate the build-up of the cross-circular polarization we present the rate equation model accounting for the valley-independent generation of B-excitons via relaxation from excited $|f\rangle$ states and valley-dependent stimulated scattering towards A:$1s$ excitons. To that end we introduce the generation rate of B-excitons, which is the same in both $\sigma^+$ and $\sigma^-$ polarizations, $G_{\mathrm B}$ and present the rate equations for the densities $N^\pm_{\mathrm B}$ of the $\sigma^\pm$-polarized B:$1s$-excitons in the form 
\begin{subequations}
\begin{align}
\frac{dN^+_{\mathrm B}}{dt} + \frac{N^+_{\mathrm B}}{\tau_B} + W (1+ N^+_{\mathrm A}) N^+_{\mathrm B} = G_B,\\
\frac{dN^-_{\mathrm B}}{dt} + \frac{N^-_{\mathrm B}}{\tau_B} + W (1+ N^-_{\mathrm A}) N^-_{\mathrm B} = G_B,
\end{align}
\end{subequations}
where $N_{\mathrm A}^\pm$ are the occupancies of $\sigma^\pm$ polarized A:$1s$-excitons, $\tau_B$ is the lifetime of $B$-excitons irrelated with relaxation towards the A-states, $W$ describes the rate of the relaxation to A:$1s$-excitonic state. This description is simplified as we neglect spin/valley relaxation of excitons and, moreover, the exciton relaxation from B- to A-state can be with multiple steps, in which case a cascaded process may be relevant~\cite{Liew:2013a}. The circular polarization degree of $B$-excitons can be expressed, in the limit $W\tau_B(1+ N^\pm_{\mathrm A}) \ll 1$ as
\begin{equation}
P_{\rm B} = \frac{N^+_{\mathrm B} - N^-_{\mathrm B}}{N^+_{\mathrm B} + N^-_{\mathrm B}} = -\frac{1}{2} W\tau_B  N_{\mathrm{A}:1s} P_{\rm A},
\end{equation}
where $P_{\rm A} = (N^+_{\mathrm A} - N^-_{\mathrm A})/N_{\mathrm{A}:1s}$ is the circular polarization of the $A$-excitons, $N_{\mathrm{A}:1s}= N^+_{\mathrm A} + N^+_{\mathrm A}$ is the total number of $A$-excitons. Clearly, the polarization $P_B$ of the upconverted B:$1s$-exciton emission is reversed as compared with $P_{\rm A}$ and it is the larger the more excitons are created by the laser, i.e. the larger $N_{\mathrm{A}:1s}$. For the incident intensity $I=100$~$\mu$W/$\mu$m$^2$ and A-exciton lifetime $\tau_A=1$~ps we have, in accordance with Eq.~\eqref{1PA} the A-exciton density $N_x=10^{10}$~cm$^{-2}$. To obtain an estimate of the occupancy of a single quantum state we present
\begin{equation}
\label{deg}
N_{\mathrm{A}:1s} = \frac{\hbar^2 N_x}{M \Delta},
\end{equation}
where $M\approx 0.5m_0$ with $m_0$ being free electron mass, is the exciton effective mass and $\Delta$ is the energy width of exciton distribution. The right hand side of Eq.~\eqref{deg} plays a role of degeneracy parameter. For fully thermalized excitons $\Delta = k_B T\approx 0.3$~meV, $N_{\mathrm{A}:1s}\approx 0.05$. However, the thermalization does not take place under resonant excitation conditions and low temperatures because the exciton-acoustic phonon scattering is relatively weak in TMD materials and strongly exceed the radiative lifetime of excitons~\cite{PhysRevB.94.205423}. The lowest limit for $\Delta$ comes from the laser linewidth, which is extremely narrow in our experiments with \emph{cw} excitation. Taking $\Delta=3\times 10^{-2}$~meV and $W\tau_B=0.5$ (evaluation of these quantities is beyond the scope of the present paper) and $P_A=100$~\% we have for $P_B=-15\%$ in the order of magnitude agreement with experiment.
